# Modulation of Memristive Characteristics by Dynamic Nanoprecipitation inside Conical Nanopores

Zhe Liu,[1,2#] Hongwen Zhang,[1,2#] Di Liu,[3] Tianyi Sui,[4] and Yinghua Qiu[1,2]*

1. Key Laboratory of High Efficiency and Clean Mechanical Manufacture of Ministry of Education, State Key Laboratory of Advanced Equipment and Technology for Metal Forming, School of Mechanical Engineering, Shandong University, Jinan, 250061, China

2. Shenzhen Research Institute of Shandong University, Shenzhen, 518000, China

3. Department of Prothodontics, School and Hospital of Stomatology, Cheeloo College of Medicine, Shandong University, Jinan, 250012, China.

4. School of Mechanical Engineering, Tianjin University, Tianjin, 300072, China

# These authors contributed equally.

*Corresponding author: yinghua.qiu@sdu.edu.cn

**ABSTRACT**

Nanofluidic memristors have demonstrated great potential for neuromorphic system applications with the advantages of low energy consumption and excellent biocompatibility. Here, an effective way is developed to regulate the memristive behavior of conical nanopores by leveraging the reversible formation and dissolution of nanoprecipitates induced by ion enrichment and depletion in nanopores under opposite voltages. Through the interplay between precipitation dynamics at the pore tip and the ion enrichment/depletion inside the nanopore, conical nanopores exhibit pronounced current hysteresis loops in the presence of $CaHPO_4$, a slightly soluble inorganic salt. The memristive characteristics are found to be strongly dependent on the concentration of $CaHPO_4$, besides the applied voltage amplitude and scan rate. Under the stimulation of pulse voltages, ionic current demonstrates stable learning and forgetting processes with robust switching stability and effective reset capability, which is similar to the short-term plasticity characteristics of biological synapses. Our research may provide a straightforward and tunable approach for the design of nanofluidic memristors.

**TOC**

A nanofluidic memristor based on conical nanopores is demonstrated, where voltage-induced ion enrichment and depletion govern the reversible formation and dissolution of CaHPO$_4$ nanoprecipitates. The resulting dynamic ionic modulation gives rise to tunable hysteresis and synaptic-like short-term plasticity. This system offers a minimalistic route toward neuromorphic iontronic functions.

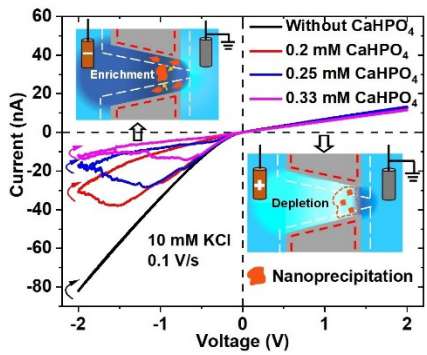

**I. Introduction**

Memristors are fundamental circuit elements capable of both processing and storing information, which can find extensive applications in various fields, such as non-volatile memory devices and neuromorphic computing.[1, 2] With advancements in nanofabrication techniques and nano-devices, various nanofluidic memristors have been developed and characterized, which achieve memristive behaviors through the dynamic regulation of ion transport through nanopores.[3, 4] Owing to their high biocompatibility and low energy consumption, nanofluidic memristors demonstrate significant potential in logic ionic circuits,[3, 5] bio-machine interfaces,[6-8] and brain-inspired neuromorphic systems.[9, 10]

The memristive characteristics of nanofluidic memristors originate from the time mismatch between ion transport and the application of external stimuli,[11, 12] which manifests as nonlinear ionic conduction and results in pinched hysteresis loops in current-voltage curves.[4] In 2011, Siwy et al.[13] first demonstrated the nanofluidic memristive behavior using hydrophobic polymer nanopores, where the wetting property of pore walls was modulated by applied electric fields. In 2012, Wang et al.[14] reported similar behavior in glass nanopipettes through control of the voltage scan rate. Later, nanofluidic memristors have been constructed using a variety of mechanisms to regulate the ion transport through nanopores,[3, 4] including ion–surface interactions,[15, 16] ion concentration polarization,[17] formation of ion enrichment/depletion inside nanopores,[18, 19] construction of a liquid/liquid interface,[20, 21] the Wien effect in sub-nanometer channels,[22, 23] and structural deformation of the nanopore/channel.[24, 25]

Due to the asymmetric geometry, conical nanopores serve as a convenient and versatile platform for investigating ion current rectification (ICR),[26] which arises from ion enrichment and depletion under opposite voltages inside the nanopore.[12, 27]

Because of the time-dependent formation of ion enrichment and depletion, conical nanopores provide a suitable platform for constructing nanofluidic memristors.[14, 19, 28] At low voltage scan rates, conical nanopores exhibit ICR behaviors. As the scan rate increases, memristive characteristics become evident. However, at a super-high scan rate, the memristive phenomenon vanishes and is replaced by pure resistive properties.[12, 29] In addition to the voltage scan rate, multiple parameters have been confirmed to regulate the memristive characteristics of conical nanopores, such as applied voltage magnitude,[19] nanopore parameters,[15, 30] solution conditions,[18, 31] surface modification with polymers,[16, 32] and the circuit configuration, i.e. series connection or parallel connection.[33]

At the open state of conical pores, both cations and anions accumulate inside the nanopore. For slightly soluble inorganic salts, this can lead to nanoprecipitates once the ionic concentration exceeds the solubility limit, resulting in nanopore blockade.[34] In the closed state, reversed voltages induce the depletion of cations and anions, which facilitates the dissolution of previously formed nanoprecipitates. Through the dynamic process of precipitation and dissolution, Powell et al.[35] observed current oscillation in conical nanopores. Similarly, Laucirica et al.[36] confirmed that the reversible formation and dissolution of nanoprecipitates directly control the ion transport inside conical nanopores. Tsutsui et al.[37] further directly observed the formation of insoluble aluminium phosphate in silicon nitride nanopores and developed in-pore chemical reaction-driven nanofluidic memristors through nanoprecipitate dynamics. Previous research mainly focused on constructing memristors utilizing nanoprecipitates, but overlooked the fundamental current regulation mechanisms and parameter-dependent control of memristive characteristics.

Considering the significant response time required for the dynamic formation and dissolution of nanoprecipitates inside conical nanopores, we propose a strategy to modulate the memristive properties of conical nanopores through this time-dependent process. This mechanism offers a straightforward approach to designing nanofluidic memristors. In this work, the regulation of ionic current was systematically investigated through the reversible formation and dissolution of nanoprecipitates inside single conical nanopores, which effectively modulate the memristive properties of conical nanopores. By leveraging nanoprecipitation dynamics at the nanopore tip, enhanced hysteresis loops were observed in the current-voltage curves. The memristive characteristics are directly related to the concentration of $CaHPO_4$, which can be regulated by adjusting the voltage amplitude and scan rate. Finally, under stimulation of periodic pulse voltages, conical nanopores were able to emulate the short-term synaptic plasticity, with current exhibiting excellent learning and forgetting behaviors, high stability, as well as reliable reset characteristics.

**II. Experimental Details**

Single conical nanopores were fabricated on 12.5 μm-thick polyethylene terephthalate (PET) films using our previously developed technique that combines needle punching and chemical etching.[38] First, tungsten needles with nanoscale tips and controlled cone angles were prepared with electrochemical etching.[39] Then, a needle was driven precisely by a motor (Injection Pump LSP02-2A, Longer Pump, China) to pierce the PET membrane and create a conical defect. Next, the film containing the conical defect was mounted between two electrolyte reservoirs. A 9 M NaOH solution was introduced on the side containing the defect to initiate chemical etching at 25 °C, while a stop solution consisting of 1 M HCOOH and 1 M KCl was added to the opposite side.[40, 41] The etching process was conducted under feedback control by continuously monitoring the ionic current through the membrane, ensuring the precise fabrication of single conical nanopores. Finally, the conical nanopore was sized based on the current-voltage (I-V) curves collected in 1 M KCl at pH 8 using a picoammeter (Keithley 6487, Keithley Instruments LLC, USA).[42]

In this work, all chemicals were used as received from Sigma Aldrich. A 10 mM KCl solution was mixed with varying concentrations of $CaHPO_4$ or $CaCl_2$ at pH 8. The pH and conductivity of each solution were measured using a pH meter (Mettler Toledo FE 28, China) and a conductivity meter (Mettler Toledo FE 38, China), respectively. The fabricated conical nanopore was positioned between two reservoirs filled with electrolyte solution. Ag/AgCl electrodes were employed to apply voltages across conical nanopores. The electrode setup is shown in Figure S1a. The ground and working electrodes were placed at the tip and base sides of conical nanopores, respectively.

Steady-state I-V curves were recorded using a pico-ammeter, Keithley 6487. Dynamic ion currents were recorded at a low sampling frequency of 1 kHz using an electrochemical workstation (CHI 760E, CH Instruments Inc., USA) under triangular (Figure S1b i) and rectangular waveform functions (Figure S1b ii). Voltage scan rates ranging from 0.1 to 20 V/s were applied during the dynamic I-V curves. For the current traces at a high sampling frequency of 10 kHz, Axopatch 200B and Digidata 1550B (Molecular Devices Inc., USA) were applied at various voltages (Figure S1b iii). Data analysis was conducted using Clampfit 11.0 (Molecular Devices Inc., USA).

Please note that conical nanopores in this work are applied to form the ion enrichment and depletion, enabling the dynamic formation and dissolution of nanoprecipitates for modulating memristive performance. There are no strict requirements for the morphology and size of the nanopores. We believe that the application of track-etched conical nanopores, as well as other nanopores of asymmetric geometries, which present strong ionic current rectification,[27, 43] can work well for the proposed strategy of constructing nanofluidic memristors with the dynamic formation and dissolution of nanoprecipitates.

**III. Results and discussion**

Charged conical nanopores with asymmetric geometries can exhibit different current values at equal voltages but opposite polarities, the phenomenon known as ICR. This effect, typically observed at low voltage scan rates, originates from the ionic selectivity of the nanopore. Our previous studies have demonstrated that the establishment of ion enrichment and depletion zones within conical nanopores is a time-dependent process.[12] Therefore, when dynamic voltage scans are applied, the corresponding ion distribution deviates from its steady-state profile, causing the memristive characteristics of conical nanopores.[14]

$CaHPO_4$ is a slightly soluble inorganic salt, with a bulk solubility product $K_{sp}$ of ~$1 \times 10^{-7}$ M at ambient conditions.[35] In the presence of $CaHPO_4$ and KCl solutions, facilitated by the ion enrichment and depletion inside conical nanopores, Powell et al.[35] investigated the dynamic formation and dissolution of $CaHPO_4$ nanoprecipitates, which required a certain amount of time to complete. From the recorded current traces, the reference current was attributed to the background solution, while the dynamic behavior of nanoprecipitation caused the observable current oscillation. Their study provided practical guidance for the design of ion oscillators.

Here, inspired by the work of Powell et al.,[35] considering the time required for the ion enrichment inside conical nanopores and the subsequent formation of nanoprecipitates, we designed a series of nanofluidic experiments to investigate the memristive characteristics of conical nanopores modulated by the dynamic formation and dissolution of nanoprecipitates. Single conical nanopores were fabricated on 12.5-μm-thick PET membranes with a typical cone angle of ~15°. After fabrication, I-V curves were obtained in 1 M KCl solution (Figure S2), and the corresponding tip and base diameters of conical nanopores were estimated based on the ionic conductance measurements. In this work, conical nanopores with tip diameters ranging from 10 to 60 nm were applied. Due to the low solubility of $CaHPO_4$, the application of $CaHPO_4$ alone results in a low signal-to-noise ratio. It is necessary to select a KCl solution of appropriate concentration as the background solution.[35]

Figure S4a presents the I-V curves of a conical PET nanopore measured in KCl solutions with concentrations varying from 1 to 200 mM. The ICR performance was evaluated using the ICR ratio calculated by $|I_{-1V}/I_{1V}|$.[44] According to previous studies, a higher ICR ratio reflects a more significant ion enrichment and depletion inside the nanopore.[43] Figure S4b exhibits the dependence of the ICR ratio on the KCl concentration, revealing an increasing-decreasing trend as the concentration varies.[45-47] The maximum ICR ratio of ~10.6 was observed at ~10 mM, which was selected as the background solution for experiments involving the addition of sub-millimolar $CaHPO_4$. Please note that as the salt concentration of the background solution increases from 10 mM to 100 mM, the hysteresis loop disappears in the I-V curves due to the better screening of surface charges by counterions (Figure S3).

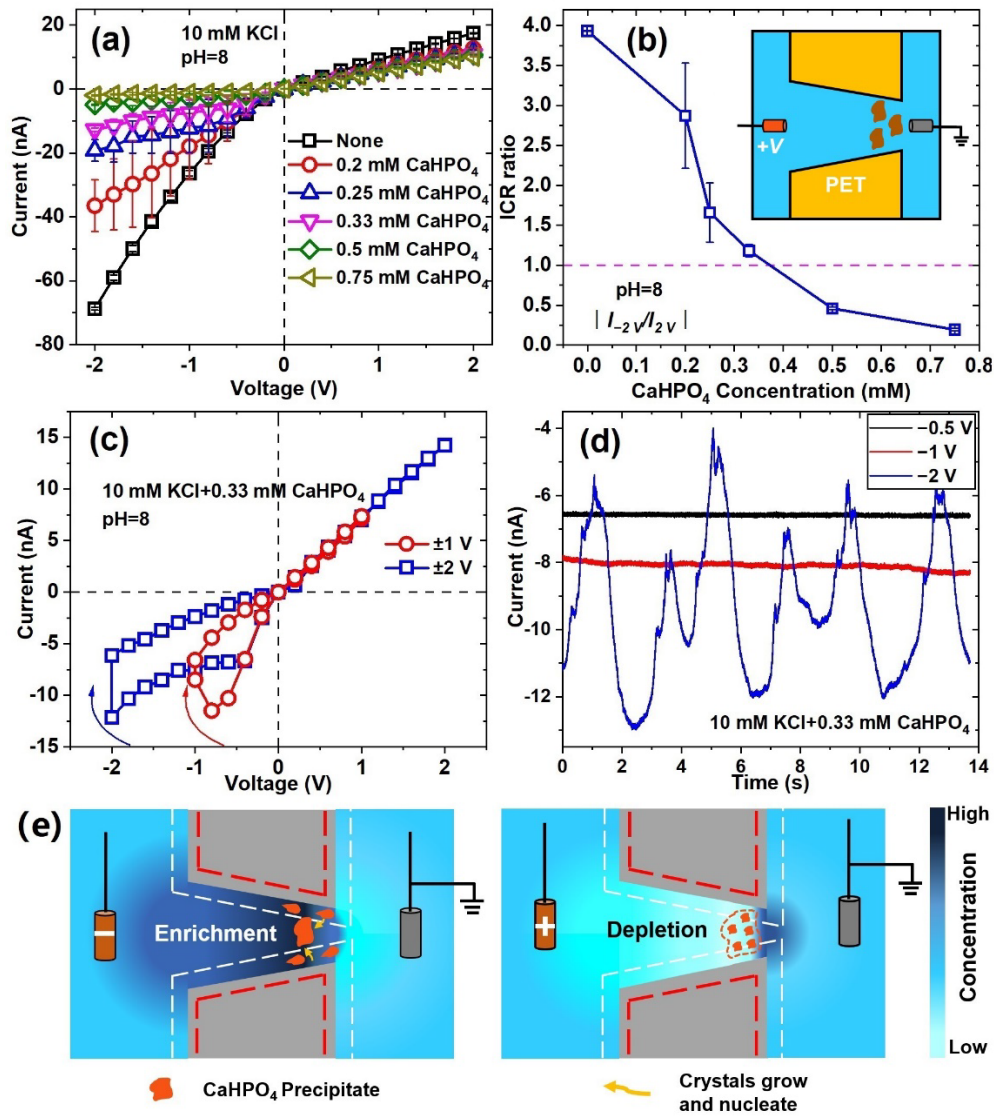

Figure 1. Characteristics of ion current in various $CaHPO_4$ concentrations. (a) I-V curves with $CaHPO_4$ concentrations varying from 0 to 0.75 mM. (b) Corresponding ICR ratios at ±2 V. The inset shows the variation of ICR ratios with the applied voltage. (c) I-V curves at different voltage amplitudes in 0.33 mM $CaHPO_4$. (d) I-t traces at different voltages in 0.33 mM $CaHPO_4$. (e) Schematic diagrams of the formation and dissolution of $CaHPO_4$ nanoprecipitates. Bars are presented as mean ± SD ($n$ = 4). The pore length is 12.5 μm, the tip diameter is 12.5 nm, and the cone angle is ~15°.

Figure 1a shows I-V curves through the conical PET nanopore in 10 mM KCl containing 0 ~0.75 mM $CaHPO_4$. The I-V curve recorded in 10 mM KCl alone presents excellent ICR characteristics with a high repeatability, indicating both the asymmetric conical geometry of the nanopore, as well as stable physical and chemical properties of pore surfaces.[26] With the addition of $CaHPO_4$ into the 10 mM KCl solution, the current at negative voltages, i.e., the open state, decreases significantly, accompanied by noticeable increases in the current fluctuation, as reflected by the large error bars. This phenomenon is attributed to the formation of $CaHPO_4$ nanoprecipitates inside conical nanopores (see below), which induces the dynamic fluctuation of the ionic current. As the bulk $CaHPO_4$ concentration increases, the current values decrease. For example, at −2 V, the current reaches ~−35 nA in 0.2 mM $CaHPO_4$, which is ~50% of the current ~−68 nA observed in the pure KCl. With 0.75 mM $CaHPO_4$, the current further decreases to ~−1.6 nA. In another experiment, additional I-V curves were recorded in mixed solutions of 10 mM KCl and varying concentrations of $CaHPO_4$ at ±1 V (Figure S5a). In this case, the ionic current decreased less in the 0.2 mM $CaHPO_4$ solution compared to the solution without $CaHPO_4$. A comparison of the error bars at ±2 V and ±1 V demonstrates that the current oscillations are more significant at a higher voltage, indicating the strongly voltage-dependent regulation of ionic current. At −1 V, the current gradually decreases with the $CaHPO_4$ concentration increasing, indicating that a sufficiently high bulk concentration of $CaHPO_4$ is required to enable the nanoprecipitation-based current modulation inside conical nanopores.

ICR ratios are analyzed from I-V curves at various bulk concentrations of $CaHPO_4$ to investigate the modulation of ionic current by nanoprecipitation (Figure 1b). As the bulk concentration of $CaHPO_4$ increases from 0 to 0.75 mM, both the magnitude and fluctuation of the ICR ratio gradually decrease. Interestingly, the ICR ratio drops below 1 when the $CaHPO_4$ concentration exceeds 0.5 mM, indicating the appearance of a reversed ICR phenomenon. This inversion originates from the decreased ionic current at the open state as $CaHPO_4$ concentration increases, while the current at the closed state remains nearly unchanged. Figure S6 shows the dependence of the ICR ratio on the applied voltage at different $CaHPO_4$ concentrations. In pure KCl solutions, the ICR ratio increases monotonously with the applied voltage, consistent with previous reports.[26, 27] After the addition of $CaHPO_4$, the ICR ratio starts to fluctuate at various voltages due to the dynamic formation and dissolution of nanoprecipitates. When the $CaHPO_4$ concentration is below 0.33 mM, the ICR ratio presents an increasing-decreasing profile with voltage. However, once the bulk concentration of $CaHPO_4$ exceeds 0.5 mM, the ICR ratio decreases continuously as the voltage increases. These results further confirm that both voltage and $CaHPO_4$ concentration are critical parameters in regulating the ionic current through conical nanopores.

The conductivity of the mixed solutions with different $CaHPO_4$ concentrations was measured, which remained constant at ~145 mS/m (Figure S5b). This indicates that the decrease in ion current at −2 V with the bulk concentration of $CaHPO_4$ increasing is not caused by the change in the solution conductivity. Additional I-V curves were collected in solutions containing 10 mM KCl and $CaCl_2$ with concentrations varying from 0.2 to 0.75 mM. From Figure S7, I-V curves obtained in different $CaCl_2$ solutions present almost the same current values, indicating that the presence of $Ca^{2+}$ ions alone does not account for the current suppression at −2 V.[48] Therefore, we conclude that the decrease in ionic current and the altered ICR ratios may be induced by the blockade effect of the nanoprecipitates inside conical nanopores.[35, 37]

Figure 1c shows I-V curves in a mixed solution of 10 mM KCl and 0.33 mM $CaHPO_4$ under ±1 V and ±2 V. Pronounced hysteresis loops are observed in the I-V curves under negative voltages, which are responsible for the large current error bars. To gain further insight into the dynamic modulation of ionic current by nanoprecipitation, current-time (I-t) traces were further investigated at constant voltages in the same solution (Figure 1d). With the voltage increasing, the amplitude of the current oscillation becomes more significant. At −0.5 V, the ionic current remains stable at ~−7 nA. When the voltage increases to −2 V, periodic oscillations between −6 nA and −12 nA emerge. Analysis of the current probability histogram and power spectrum reveals that the oscillatory behaviors of ionic currents positively correlate with the applied voltage (Figure S8).[35] This provides direct evidence that the observed current fluctuations originate from the voltage-dependent formation and dissolution of $CaHPO_4$ nanoprecipitates in the conical nanopores. Specifically, the current fluctuation under negative voltages originates from the continuous growth and dynamic movement of nanoprecipitates inside the confined geometry of conical nanopores. Compared to the work by Powell et al.,[13] our system exhibits a lower current oscillation frequency. This may result from the large conical nanopores applied in this work, which reduce the ion enrichment efficacy (Figure S9) and consequently slow nanoprecipitate formation kinetics.

Figure 1e shows the dynamic formation and dissolution of nanoprecipitates in response to opposite voltages. At negative voltages, ions are enriched inside conical nanopores. Considering the high ion concentration inside the electric double layers (EDLs) near the pore tip,[12] nanoprecipitate initially forms in the EDLs of this region. As the ion enrichment continues, the increasing amount and size of nanoprecipitates lead to partial or complete pore blockage, resulting in a marked decrease in current. Under positive voltages, ion depletion occurs inside conical nanopores. When the $CaHPO_4$ concentration drops below its solubility constant, nanoprecipitates begin to dissolve gradually, which results in the reopening of the nanopore. From Figure S5b, the mixed solutions containing different concentrations of $CaHPO_4$ and 10 mM KCl exhibit a similar electrical conductivity. Consequently, the current data under the positive voltage remain nearly identical regardless of the presence or absence of $CaHPO_4$.

With the mixed solutions of CaHPO$_4$ and KCl, I-V curves in conical nanopores present pronounced hysteresis loops, which are closely related to the dynamic formation, growth, and dissolution of CaHPO$_4$ nanoprecipitates. Therefore, revealing the dynamic regulation of ionic current by nanoprecipitation provides a foundation for enhancing the memristive properties of conical nanopores utilizing the dynamic changes of nanoprecipitates.

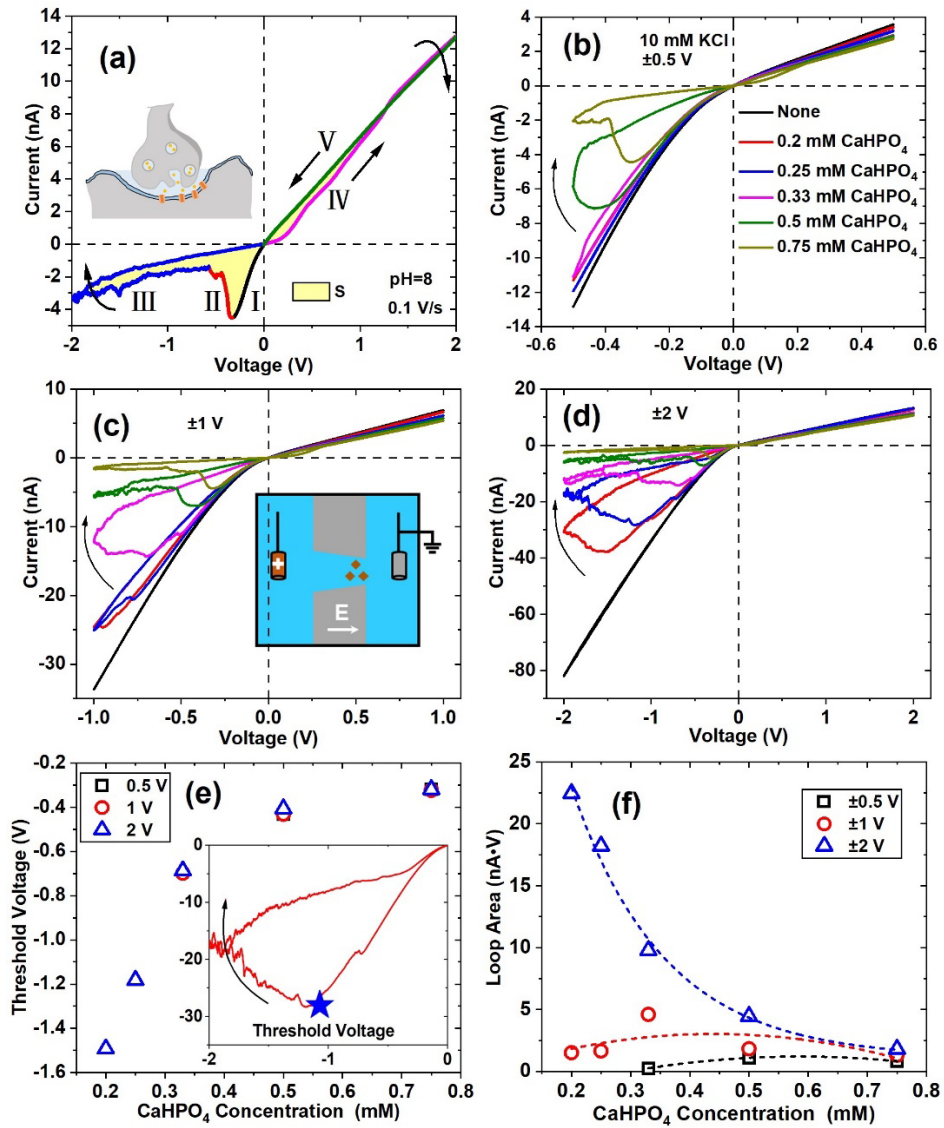

Figure 2. Influence of the voltage on memristor characteristics in various CaHPO$_4$ concentrations. (a) A characteristic hysteresis loop. Arrows represent the scan direction

of the voltage. (b-d) I-V curves at various CaHPO$_4$ concentrations under ± 0.5 V (b), ±1 V (c), and ±2 V (d). (e) Dependence of the threshold voltage on CaHPO$_4$ concentration. The inset shows the definition of the threshold voltage. (f) Loop area enclosed by the I-V curve with various CaHPO$_4$ concentrations at different voltages.

Figure 2a presents an I-V curve with the hysteresis loop in conical nanopores with the CaHPO$_4$ solution. The loop can be divided into five parts, each corresponding to a distinct nanoprecipitation state inside the nanopore. At the open state, the ionic current initially increases and then decreases, followed by stabilization with the voltage increasing, exhibiting the behavior of the negative differential resistance (NDR).[35, 49, 50] Part I displays a linear current increase with the voltage due to the ion accumulation inside the nanopore. When CaHPO$_4$ concentration exceeds its solubility limit, nanoprecipitates abruptly form and result in a sharp current drop (Part II). The presence of nanoprecipitates inhibits ion transport, leading to reduced current during the voltage scan from −2 to 0 V compared to the reverse scan from 0 to −2 V (Part III). As the voltage sweeps from negative to positive, the system switches to the closed state. The nanoprecipitates start to dissolve because of the enhanced ion depletion at increasingly positive voltages (Part IV). The current in Part IV is smaller than that in Part V, where the voltage scans from 2 to 0 V and the nanoprecipitates are totally dissolved. At the end of Part V, the ICR properties in the conical nanopore are restored. Please note that Parts IV and V intersect at ~1.25 V, marking the threshold voltage for the full dissolution of nanoprecipitates. The hysteresis loop contains the dynamic formation, growth, and dissolution of CaHPO$_4$ nanoprecipitates, demonstrating memristive behaviors with the reset capability.

Figure 2b shows the I-V curves measured at ±0.5 V under various bulk concentrations of CaHPO$_4$. Here, we mainly focus on the ionic current behaviors at negative voltages, where nanoprecipitation is expected to occur. In solutions containing 0, 0.2, and 0.25 mM $CaHPO_4$, significant ICR phenomena are observed. The current values with $CaHPO_4$ remain slightly lower than those without $CaHPO_4$. This may be due to the $Ca^{2+}$-ion-induced partial screening of surface charges on nanopore walls.[51] With the bulk concentration of $CaHPO_4$ increasing from 0.33 to 0.75 mM, the hysteresis loops in I-V curves exhibit various shapes. Because of the low applied voltages, at 0.33 mM $CaHPO_4$, the limited accumulation of nanoprecipitates is insufficient to fully block the nanopore tip. As the $CaHPO_4$ concentration increases to 0.5 mM, NDR appears in the I-V curve. In 0.75 mM $CaHPO_4$, the characteristic hysteresis loop in the I-V curve is formed. As the applied voltage amplitude increases to ±1 and ±2 V (Figures 2c and 2d), both NDR and characteristic hysteresis loops appear at a lower bulk concentration of $CaHPO_4$. At ±2V, the hysteresis loop becomes obvious even at 0.2 mM $CaHPO_4$ (Figure 2d). These results demonstrate that both the bulk concentration of $CaHPO_4$ and applied potential play critical roles in modulating the shape of the hysteresis loops.

The threshold voltage and loop area are analyzed to quantitatively evaluate the characteristics of the hysteresis loops. The threshold voltage is defined as the voltage where the NDR first appears, denoted by the star symbol shown in the inset of Figure 2e. From Figure 2e, the threshold voltage decreases as the bulk concentration of $CaHPO_4$ increases. Under a higher concentration of $CaHPO_4$, a lower threshold voltage is obtained. As the $CaHPO_4$ concentration decreases from 0.75 to 0.2 mM, the threshold voltage increases from ~−0.32 to ~−1.5 V. This is attributed to that a higher $CaHPO_4$ concentration in bulk solutions facilitates the faster local accumulation of ions, thereby reaching the nucleation threshold with a lower applied voltage. Please note that under various voltage amplitudes, the obtained threshold voltage remains constant.

The loop area, defined as the area enclosed by the I-V curve, can be used to characterize the memristive performance of conical nanopores.[15, 52] As highlighted by the yellow region in Figure 2a, a larger loop area corresponds to a stronger memristive behavior. Figure 2f presents the dependence of the hysteresis loop area on the applied concentration of $CaHPO_4$ under different voltages. At higher voltages, the loop areas are significantly larger than those observed at lower voltages. At ±0.5 V and ±1 V, the loop area exhibits an increasing-decreasing profile with the $CaHPO_4$ concentration increasing, although the overall values remain small. At low voltages and low $CaHPO_4$ concentrations, nanoprecipitates cannot form, resulting in the absence of a hysteresis loop. With the increase of the $CaHPO_4$ concentration, the formation of nanoprecipitates induces a hysteresis loop in the I-V curve. However, as the concentration of $CaHPO_4$ increases further, the threshold voltage becomes smaller, resulting in a decrease in the hysteresis loop area. At ±2 V, as the $CaHPO_4$ concentration increases from 0.2 to 0.75 mM, the loop area has a decreasing trend. This is due to the smaller threshold voltage for the formation of nanoprecipitates inside conical nanopores under these high voltages. The memristive effect in a conical nanopore is governed by the interplay between the $CaHPO_4$ concentration and the applied voltage.

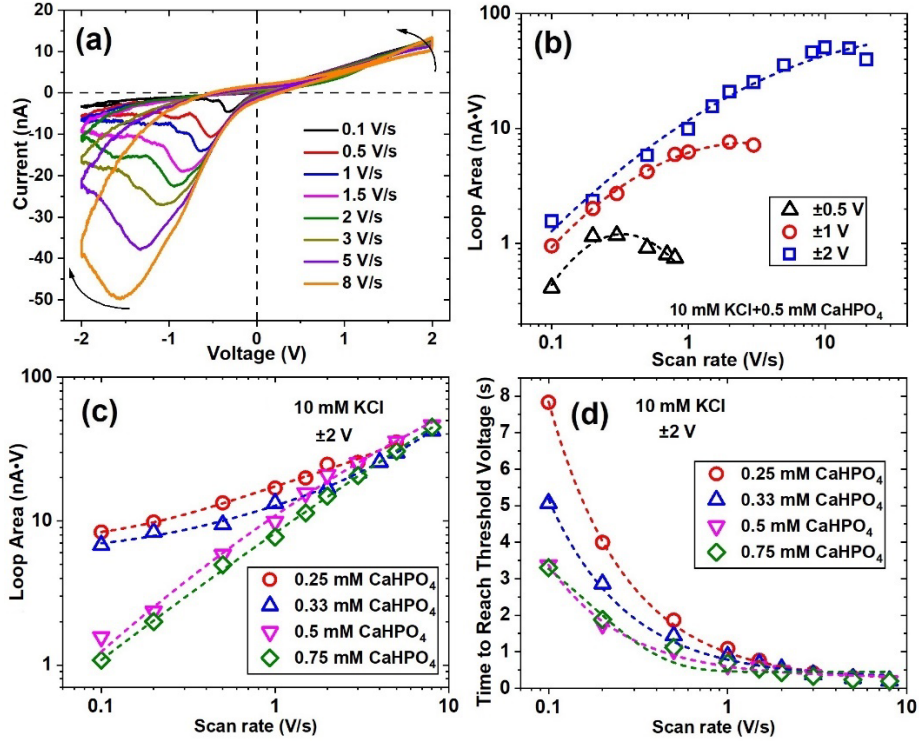

Figure 3. Influence of the voltage scan rate on memristor characteristics. (a) I-V curves at various voltage scan rates in mixed solutions of 10 mM KCl and 0.5 mM $CaHPO_4$. (b) Loop area varies with the voltage scan rate under various voltage amplitudes. (c) Loop area varies with the voltage scan rate under various $CaHPO_4$ concentrations. (d) Dependence of time to reach the threshold voltage on the voltage scan rate under various $CaHPO_4$ concentrations. Here, the time to reach the threshold voltage ($t$) is calculated by $t=V_t/v$. $V_t$ and $v$ represent the threshold voltage and voltage scan rate in the I-V curve.

Considering the required time for the ionic enrichment inside conical nanopores, we investigated how the voltage scan rate modulates the memristor characteristics, aiming to establish guidelines for nanofluidic memristor optimization and applications.[3, 4] Here, the voltage scan rates ranging from 0.1 to 20 V/s are applied.

Figure 3a shows the I-V curves at different voltage scan rates ranging from 0.1 to 8 V/s in 0.5 mM $CaHPO_4$. The loop area of I-V curves increases monotonously with the voltage scan rates. Under steady-state conditions (0.1 V/s), the current remains nearly constant across the voltage range due to the sustained nanopore blockage by $CaHPO_4$ nanoprecipitates, resulting in weakened memristive features. As the voltage scan rate increases, ions are prevented from reaching steady-state enrichment inside the conical nanopore due to the longer time required for the ion response than the variation of applied voltages. The delayed equilibration of ion states extends the formation, growth, and dissolution processes of nanoprecipitation. Consequently, the applied voltage for the formation/dissolution of nanoprecipitates exhibits a strong correlation with the voltage scan rate. The threshold voltage increases from −0.3 to −1.6 V as the scan rate increases from 0.1 to 8 V/s (Figure S10). Please note that through the comparison of I-V curves at different bulk concentrations of $CaHPO_4$, with the voltage scan rate increasing, a higher bulk concentration is required to achieve similar memristive behaviors of conical nanopores (Figure S11).

To quantitatively analyze the memristive behavior of conical nanopores under different voltage scan rates, Figure 3b plots the dependence of the loop area enclosed by the I-V curve on the scan rate under different applied voltage amplitudes.[52] For the three considered amplitudes, the hysteresis loop areas present an increasing-decreasing profile as the voltage scan rate increases. Both the maximum loop area and the corresponding optimal scan rate are strongly influenced by the applied voltage amplitude. Under ±0.5 V, the loop area peaks at ~1.2 nA·V at the voltage scan rate of ~0.3 V/s. While at ±2 V, the loop area reaches a much higher maximum of ~50.7 nA·V, which requires a significantly faster scan rate of ~10 V/s. Therefore, the memristive performance of conical nanopores can be modulated by reasonably adjusting either the scan rate or the applied voltage amplitude.

Figure 3c exhibits the hysteresis loop areas in conical nanopores under various bulk concentrations of $CaHPO_4$. The voltage is fixed at ±2 V, and the voltage scan rate varies from 0.1 to 10 V/s. At lower scan rates from 0.1 to 2 V/s, the obtained loop area presents a negative correlation with the bulk concentration of $CaHPO_4$. This behavior is primarily due to the reduced ion accumulation efficiency at lower bulk concentrations, which delays the formation of nanoprecipitates and weakens current blockade effects. At low concentrations, more time and a stronger electric field are required to achieve the critical ion concentration necessary for nanoprecipitation inside the nanopore. A critical threshold of the bulk concentration of $CaHPO_4$ is required to achieve the formation of nanoprecipitates since nanoprecipitation fails to form at a $CaHPO_4$ concentration lower than 0.33 mM. For the voltage scan rate above 5 V/s, the loop area becomes independent of the applied $CaHPO_4$ concentration (Figures S11 and S12). As the voltage scan rate further increases to ~20 V/s, because ions do not have enough time to respond to rapidly changing electric fields, both ICR and memristive characteristics (Figure S13) are eliminated in conical nanopores.[29]

In this work, the slow process of nanoprecipitate formation/dissolution is applied to enhance the memristive behavior in conical nanopores. Figure 3d provides the time required to reach the threshold voltage where the precipitation begins to form. Under various bulk concentrations of $CaHPO_4$ ranging from 0.25 to 0.75 mM, the time to reach the threshold voltage decreases as the voltage scan rate increases. In 0.25 mM $CaHPO_4$, the response time to reach the threshold voltage decreases from ~7.8 s at 0.1 V/s to ~0.3 s at 5 V/s, corresponding to ~26 times faster. This pronounced decrease is attributed to the significantly lower ion flux required to reach the critical nanoprecipitation concentration under high scan rates than that under steady-state conditions (Figure S14). Here, the ion flux is calculated as the amount of electric charge ($Q$) required for the formation of nanoprecipitates by $Q=I \cdot t$, with $t=V_t/v$. $V_t$, $v$, and $I$ represent the threshold voltage, voltage scan rate, and the current corresponding to the threshold voltage in the I-V curve, respectively. As the bulk concentration of $CaHPO_4$ increases, the critical concentration required for nanoprecipitation can be achieved in a shorter time. However, under high voltage scan rates above ~2 V/s, the time required for the nanoprecipitate formation becomes nearly independent of the bulk concentration of $CaHPO_4$ (Figure S14). In this case, the reduced ionic migration rates and fluxes by faster nanoprecipitate formation enhance the memristor characteristics in conical nanopores.

In addition, the growth time of nanoprecipitates under various voltage scan rates is obtained, as shown in Figure S15. The growth time is defined as the duration of decreased current under enhanced voltages (Figure S16a). The growth time of nanoprecipitates increases with the voltage scan rates, which is attributed to the reduced ion fluxes at higher scan rates, thereby limiting both the size and quantity of nanoprecipitates (Figure S16b). Furthermore, increasing the bulk $CaHPO_4$ concentration accelerates the growth process of nanoprecipitates, which enhances the supply of ions inside conical nanopores. Considering that the formation process of nanoprecipitates is modulated by adjusting the bulk $CaHPO_4$ concentration, voltage amplitude, and voltage scan rate, precise control can be achieved in the memristive characteristics of the conical nanopore.

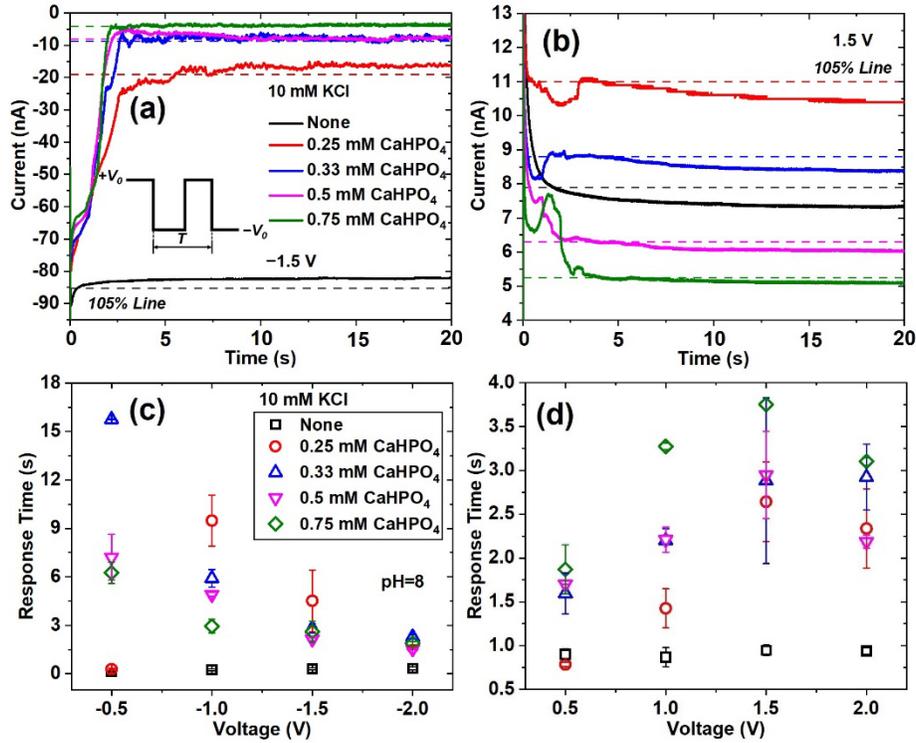

Figure 4. Dynamic current response modulated by the nanoprecipitation. (a-b) Current-time traces under various $CaHPO_4$ concentrations at −1.5 V (a) and 1.5 V (b). (c-d) Current response times under various $CaHPO_4$ concentrations at negative (c) and positive (d) voltages. Bars are presented as mean ± SD ($n$ = 3).

The dynamic response characteristics of nanofluidic memristors were investigated under the step voltage stimuli,[12] using bidirectional pulses composed of negative and positive voltage segments with a 1:1 duration ratio. Each pulse cycle lasted 40 s to ensure sufficient durations for the learning and forgetting processes.[16, 52] Figure S17 shows the current-time traces recorded during five successive bidirectional pulses. Under a constant voltage and $CaHPO_4$ concentration, the current exhibits high reproducibility in shape and amplitude in five consecutive bidirectional pulses. It demonstrates that ionic current can achieve equilibrium within 20 s, which exhibits remarkable temporal stability and reset characteristics. Based on our previous simulation work,[12] the response time of the conical nanopore is defined as the duration required for the current to reach a value within 5% deviation from its steady-state at 20 s. This time is denoted as the learning or forgetting time of the nanofluidic memristor (Figure S18). Extended testing with 60-second cycles (Figure S19) maintained the excellent repeatability during the prolonged operation. These results prove our nanofluidic memristor offers exceptional stability and switching durability, highlighting its potential for neural network applications.

Figure 4a presents the current-time traces following a voltage step from 1.5 to −1.5 V at various $CaHPO_4$ concentrations. At −1.5 V, in the case without $CaHPO_4$, the ionic current rapidly stabilizes within ~0.2 s. In contrast, in solutions with $CaHPO_4$, the ionic current through nanopores undergoes a sharp initial decline followed by progressive stabilization. The characteristics reflect the formation and growth of nanoprecipitates, which induce current blockades in the nanopore. Notably, as the bulk concentration of $CaHPO_4$ increases, the corresponding ionic current gets significantly suppressed.

The ratios of the steady-state current to the corresponding initial current were analyzed across different bulk concentrations of $CaHPO_4$ (Figure S20), which exhibit a decreasing trend with the increase of the applied voltage. Specifically, in the solution with 0.5 mM $CaHPO_4$, the steady-state current accounts for ~63.15% of the corresponding initial current at −0.5 V. While the steady-state current decreases to ~3.87% of the initial current at −2 V, corresponding to a ~16-time decrease than that at −0.5 V. This decline is attributed to the increased volume and quantity of nanoprecipitates formed under higher voltages, which intensify the blockade of ionic transport through the nanopore. Under a constant applied voltage, the current ratio decreases with the increase of the bulk concentration of $CaHPO_4$ (Figure S20). At higher $CaHPO_4$ concentrations, a more significant current blockade appears in the nanopore. At −1.5 V, the steady-state current of 0.33 mM $CaHPO_4$ solution accounted for ~61.72% of the initial current, while that of the 0.75 mM CaHPO$_4$ solution is only ~1.19%, corresponding to a ~52-fold reduction.

Figure 4b shows the current-time traces under a step voltage from −1.5 to 1.5 V at various CaHPO$_4$ concentrations. In the case without CaHPO$_4$, the current gradually decreases and stabilizes. However, in the mixed solutions with CaHPO$_4$, the current first exhibits a distinct decreasing-increasing-decreasing profile, followed by slow stabilization. The increase in current results from the dissolution of nanoprecipitates formed under negative voltages, which progressively mitigates the blockade effect as the voltage reverses from negative to positive. The subsequent decrease in current arises from the enhancement in the ion depletion inside conical nanopores, particularly near the tip region. Ion depletion is constrained by the bulk concentration of KCl, leading to comparable steady-state current values in solutions with and without CaHPO$_4$.

Figures 4c and 4d plot the response time extracted from the current traces under step voltages. In solutions without CaHPO$_4$, the response time remains nearly constant across the tested voltage range, showing minimal dependence on the voltage values at both positive (~0.25 s) and negative (~0.14 s) biases. At negative voltages, the response time for solutions with CaHPO$_4$ decreases significantly with the increase of the voltage. In 0.33 mM CaHPO$_4$, the response time sharply decreases from ~15.7 s at −0.5 V to ~2.3 s at −2 V. This trend is attributed to the accelerated ion migration by the enhanced electric field strength, which enables the faster accumulation of ions inside the nanopore. In addition, at constant voltages, the response time decreases with increasing the bulk concentration of CaHPO$_4$. At −1 V, the response time is ~9.5 s in 0.25 mM CaHPO$_4$, which is ~3.3 times higher than ~2.9 s in 0.75 mM CaHPO$_4$. This corresponds to the faster formation and growth of nanoprecipitates under the higher bulk concentration of CaHPO$_4$. From Figure 4, with the combination of the ion enrichment and dynamic formation of nanoprecipitates, the nanopore presents a substantially longer response time than that in the case without $CaHPO_4$. Note that at −0.5 V, where the applied field is insufficient to trigger nanoprecipitation, the response time in 0.25 mM $CaHPO_4$ is nearly identical to that without $CaHPO_4$.

Under the voltage reversal from negative to positive bias, additional response time is needed to dissolve previously formed nanoprecipitates and eventually reach the final equilibrium. At positive voltages, the response time presents an increasing-decreasing profile with the applied voltage. In the range from 0.5 V to 1.5 V, the response time increases linearly with the voltage.[12] Due to the larger and more nanoprecipitates formed inside conical nanopores under high negative voltages, the dissolution of these precipitates takes a longer time after switching to positive voltages. Under considered concentrations of $CaHPO_4$, the longest response time typically occurs at ~1.5 V. In 0.75 mM $CaHPO_4$, the response time increases from ~1.8 s at 0.5 V to ~3.7 s at 1.5 V, and then decreases slightly to ~3.1 s at 2 V. The response time decreases at higher voltages, due to the accelerated dissolution of nanoprecipitates by a stronger electric field strength. The response time positively correlates to the bulk concentration of $CaHPO_4$. At 1 V, the response time increases from ~1.4 s in 0.25 mM $CaHPO_4$ to ~3.3 s in 0.75 mM $CaHPO_4$, representing a ~2.4-fold increase. This is due to the large volume of nanoprecipitates formed under negative voltages, requiring longer dissolution times at higher concentrations. Here, both the strong voltage and high $CaHPO_4$ concentration can effectively prolong the forgetting time of stored information. While a further increase in the voltage can also accelerate the forgetting speed.

From Figure 4, a reasonable combination of the voltage amplitude at positive/negative bias and the concentration of $CaHPO_4$ can effectively improve the learning/forgetting ability of conical nanopores.

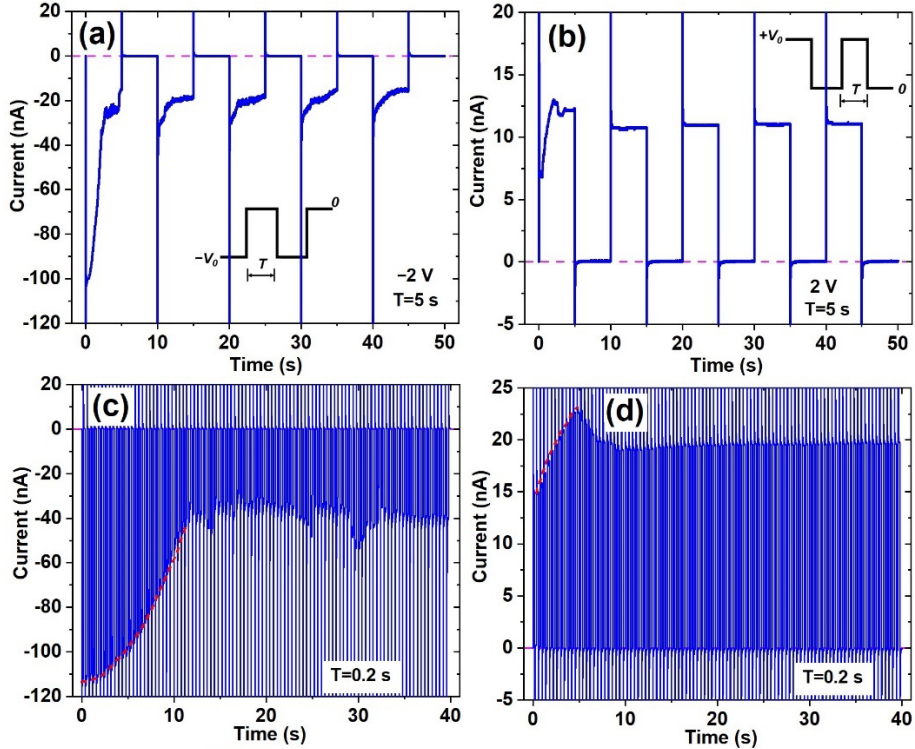

Figure 5. Current responses under a voltage pulse train (*V*= ±2 V) with varying frequencies. (a) T=5 s, *V*= −2 V. (b) T=5 s, *V*= 2 V. (c) T=0.2 s, *V*= −2 V. (d) T=0.2 s, *V*= 2 V.

Similar to the information storage function of biological synapses, nanofluidic memristors can exhibit stimulus-dependent memory behaviors, with memory retention spanning from milliseconds to several seconds. This behavior underlies synaptic plasticity, often described by the Hebbian learning rule.[16, 22] It is necessary to verify the synaptic function of the conical nanopore under the application of voltage pulses, which proves its potential application in neural computing networks. Here, we applied voltage pulses to the nanofluidic memristor, whose performance is enhanced by nanoprecipitate-mediated ionic modulation. This experimental design aims to validate their synapse-like functions and to confirm their potential in neuromorphic computing.

Paired voltage pulses are applied with negative and positive polarities, corresponding to learning and forgetting processes in the information storage.[18] The pulse interval T ranges from 5 s to 0.1 s (Figure S21). Figure 5a shows the I-t trace under pulses changing from 0 to −2 V with T=5 s in 0.5 mM $CaHPO_4$. Under negative voltages, the current variation represents the memory process. During the first voltage pulse, the ionic current decreases rapidly, followed by an almost constant current under subsequent voltage pulses. It indicates a shorter current response process under prolonged stimulus intervals.

Under voltage pulses changing from 0 to 2 V, the ionic current initially increases, followed by a gradual decrease in the stabilization (Figure 5b). The current variation represents the forgetting process under positive voltages, where the current stabilizes during the first voltage pulse. Please note that before applying the first pulse, a pre-conditioning step at −2 V for 5 s is required to ensure the nanoprecipitate formation inside the conical nanopore. Although the periodic pulses with an interval of T=5 s generate distinguishable current increases and decreases, no synaptic-like plasticity emerges under these stimuli.

Figure 5c shows the current response under voltage pulses with T=0.2 s, where the ionic current exhibits a progressive decrease across pulses, stabilizing after ~30 pulses. This behavior is consistent with the signature of the synaptic short-term plasticity. At the open state of conical nanopores, the value of steady-state currents decreases as the time interval decreases. This is primarily due to the weakened current blockade at the open state induced by the suppressed growth of nanoprecipitates under ultra-short pulses (Figure S22). Please note that the current oscillation in the response under the voltage pulse train may be induced by the stochastic processes of the formation and dissolution of nanoprecipitates. At the closed state, the current traces exhibit excellent plasticity, stabilizing after ~14 pulses (Figure 5d). The gradual current decay from ~22.8 to ~19.2 nA is attributed to the sustained ion depletion following the dissolution of nanoprecipitates, which reduces the ionic conductivity inside the conical nanopore. Notably, in this case, the stable current exceeds the steady-state current value under pulses of T = 5 s, because the pulse stimulation at a higher frequency reduces the extent of ion depletion in conical nanopores (Figure S22).

## IV. Conclusions

In this work, we developed a nanofluidic memristor based on the reversible formation and dissolution of nanoprecipitates in conical nanopores. Experimental results demonstrate that the dynamic nanoprecipitation process is the primary origin of hysteresis loops observed in I-V curves, which are directly affected by the applied $CaHPO_4$ concentration. With the increase of the voltage scan rate, the memristive characteristics of conical nanopores become more significant, conforming to the frequency-dependent characteristics of memristors. The performance of nanofluidic memristors is quantified by the hysteresis loop area in the I-V curves, which exhibit an increasing-decreasing profile with respect to the voltage scan rate. Synaptic short-term plasticity is successfully emulated through the paired-pulse voltage stimulation. Our nanofluidic memristors exhibit excellent learning and forgetting functions, along with excellent stability and reset characteristics. This study introduces a robust and tunable strategy for the construction and design of nanofluidic memristors, promoting the development of neuromorphic computing.

**SUPPLEMENTARY MATERIAL**

Strategy for the voltage application, characterization of conical nanopore sizes, I-V curves at various voltage scan rates, I-t curves at periodic pulsed voltages, and additional experiment results are in the supporting material.

**Acknowledgments**

This work was supported by the Natural Science Foundation of Shandong Province (ZR2024ME176), the National Natural Science Foundation of China (52105579), the Basic and Applied Basic Research Foundation of Guangdong Province (2025A1515010126), and the Shenzhen Science and Technology Program (JCYJ20240813101159005), the Innovation Capability Enhancement Project of Technology-based Small and Medium-sized Enterprises of Shandong Province (2024TSGC0866).

**AUTHOR DECLARATIONS**

**Conflict of Interest**

The author has no conflicts to disclose.

**DATA AVAILABILITY**

The data that support the findings of this study are available from the corresponding author upon reasonable request.

**References**

1. Mehonic, A.; Kenyon, A. J., *Nature* **2022,** *604* (7905), 255-260. DOI 10.1038/s41586-021-04362-w.
2. van de Burgt, Y.; Lubberman, E.; Fuller, E. J.; Keene, S. T.; Faria, G. C.; Agarwal, S.; Marinella, M. J.; Alec Talin, A.; Salleo, A., *Nat. Mater.* **2017,** *16* (4), 414-418. DOI 10.1038/nmat4856.
3. Xu, G.; Zhang, M.; Mei, T.; Liu, W.; Wang, L.; Xiao, K., *ACS Nano* **2024,** *18* (30), 19423-19442. DOI 10.1021/acsnano.4c06467.
4. Noy, A.; Li, Z.; Darling, S. B., *Nano Today* **2023,** *53*, 102043. DOI 10.1016/j.nantod.2023.102043.
5. Mei, T.; Liu, W.; Xu, G.; Chen, Y.; Wu, M.; Wang, L.; Xiao, K., *ACS Nano* **2024,** *18* (6), 4624-4650. DOI 10.1021/acsnano.3c06190.
6. Keene, S. T.; Lubrano, C.; Kazemzadeh, S.; Melianas, A.; Tuchman, Y.; Polino, G.; Scognamiglio, P.; Cinà, L.; Salleo, A.; van de Burgt, Y.; Santoro, F., *Nat. Mater.* **2020,** *19* (9), 969-973. DOI 10.1038/s41563-020-0703-y.
7. Wang, T.; Wang, M.; Wang, J.; Yang, L.; Ren, X.; Song, G.; Chen, S.; Yuan, Y.; Liu, R.; Pan, L.; Li, Z.; Leow, W. R.; Luo, Y.; Ji, S.; Cui, Z.; He, K.; Zhang, F.; Lv, F.; Tian, Y.; Cai, K.; Yang, B.; Niu, J.; Zou, H.; Liu, S.; Xu, G.; Fan, X.; Hu, B.; Loh, X. J.; Wang, L.; Chen, X., *Nat. Electron.* **2022,** *5* (9), 586-595. DOI 10.1038/s41928-022-00803-0.
8. Tzouvadaki, I.; Gkoupidenis, P.; Vassanelli, S.; Wang, S.; Prodromakis, T., *Adv. Mater.* **2023,** *35* (32), 2210035. DOI 10.1002/adma.202210035.
9. Kumar, S.; Wang, X.; Strachan, J. P.; Yang, Y.; Lu, W. D., *Nat. Rev. Mater.* **2022,** *7* (7), 575-591. DOI 10.1038/s41578-022-00434-z.

10. Yu, L.; Li, X.; Luo, C.; Lei, Z.; Wang, Y.; Hou, Y.; Wang, M.; Hou, X., *Nano Res.* **2024,** *17* (2), 503-514. DOI 10.1007/s12274-023-5900-y.

11. Guerrette, J. P.; Zhang, B., *J. Am. Chem. Soc.* **2010,** *132* (48), 17088-17091. DOI 10.1021/ja1086497.

12. Liu, Z.; Ma, L.; Zhang, H.; Zhuang, J.; Man, J.; Siwy, Z. S.; Qiu, Y., *ACS Appl. Mater. Interfaces* **2024,** *16* (23), 30496-30505. DOI 10.1021/acsami.4c02078.

13. Powell, M. R.; Cleary, L.; Davenport, M.; Shea, K. J.; Siwy, Z. S., *Nat. Nanotechnol.* **2011,** *6* (12), 798-802. DOI 10.1038/nnano.2011.189.

14. Wang, D.; Kvetny, M.; Liu, J.; Brown, W.; Li, Y.; Wang, G., *J. Am. Chem. Soc.* **2012,** *134* (8), 3651-3654. DOI 10.1021/ja211142e.

15. Sheng, Q.; Xie, Y.; Li, J.; Wang, X.; Xue, J., *ChemComm* **2017,** *53* (45), 6125-6127. DOI 10.1039/c7cc01047h.

16. Xiong, T.; Li, C.; He, X.; Xie, B.; Zong, J.; Jiang, Y.; Ma, W.; Wu, F.; Fei, J.; Yu, P.; Mao, L., *Science* **2023,** *379* (6628), 156-161. DOI doi:10.1126/science.adc9150.

17. Bu, Y.; Ahmed, Z.; Yobas, L., *Analyst* **2019,** *144* (24), 7168-7172. DOI 10.1039/c9an01561b.

18. Ramirez, P.; Portillo, S.; Cervera, J.; Nasir, S.; Ali, M.; Ensinger, W.; Mafe, S., *J. Chem. Phys.* **2024,** *160* (4), 044701. DOI 10.1063/5.0188940.

19. Ramirez, P.; Gómez, V.; Cervera, J.; Mafe, S.; Bisquert, J., *J. Phys. Chem. Lett.* **2023,** *14* (49), 10930-10934. DOI 10.1021/acs.jpclett.3c02796.

20. Li, P.; Liu, J.; Yuan, J.-H.; Guo, Y.; Wang, S.; Zhang, P.; Wang, W., *Nano Lett.* **2024,** *24* (20), 6192-6200. DOI 10.1021/acs.nanolett.3c05079.

21. Chen, K.; Tsutsui, M.; Zhuge, F.; Zhou, Y.; Fu, Y.; He, Y.; Miao, X., *Adv. Electron. Mater.* **2021,** *7* (4), 2000848. DOI 10.1002/aelm.202000848.

22. Robin, P.; Emmerich, T.; Ismail, A.; Niguès, A.; You, Y.; Nam, G. H.; Keerthi, A.; Siria, A.; Geim, A. K.; Radha, B.; Bocquet, L., *Science* **2023,** *379* (6628), 161-167. DOI 10.1126/science.adc9931.

23. Feng, J.; Liu, K.; Graf, M.; Dumcenco, D.; Kis, A.; Di Ventra, M.; Radenovic, A., *Nat. Mater.* **2016,** (15), 850-855. DOI 10.1038/nmat4607.

24. Zhou, X.; Zong, Y.; Wang, Y.; Sun, M.; Shi, D.; Wang, W.; Du, G.; Xie, Y., *Natl. Sci. Rev.* **2023,** *11* (4). DOI 10.1093/nsr/nwad216.

25. Emmerich, T.; Teng, Y.; Ronceray, N.; Lopriore, E.; Chiesa, R.; Chernev, A.; Artemov, V.; Di Ventra, M.; Kis, A.; Radenovic, A., *Nat. Electron.* **2024,** *7* (4), 271-278. DOI 10.1038/s41928-024-01137-9.

26. Siwy, Z. S., *Adv. Funct. Mater.* **2006,** *16* (6), 735-746. DOI 10.1002/adfm.200500471.

27. Ma, L.; Li, Z.; Yuan, Z.; Huang, C.; Siwy, Z. S.; Qiu, Y., *Anal. Chem.* **2020,** *92* (24), 16188-16196. DOI 10.1021/acs.analchem.0c03989.

28. Yang, R.; Balogun, Y.; Ake, S.; Baram, D.; Brown, W.; Wang, G., *J. Am. Chem. Soc.* **2024,** *146* (19), 13183-13190. DOI 10.1021/jacs.4c00922.

29. Wang, W.; Liang, Y.; Ma, Y.; Shi, D.; Xie, Y., *J. Phys. Chem. Lett.* **2024,** *15* (26), 6852-6858. DOI 10.1021/acs.jpclett.4c00488.

30. Wang, D.; Brown, W.; Li, Y.; Kvetny, M.; Liu, J.; Wang, G., *Anal. Chem.* **2017,** *89* (21), 11811-11817. DOI 10.1021/acs.analchem.7b03477.

31. Portillo, S.; Manzanares, J. A.; Ramirez, P.; Bisquert, J.; Mafe, S.; Cervera, J., *J. Phys. Chem. Lett.* **2024,** *15* (30), 7793-7798. DOI 10.1021/acs.jpclett.4c01610.

32. Buchsbaum, S. F.; Nguyen, G.; Howorka, S.; Siwy, Z. S., *J. Am. Chem. Soc.* **2014,** *136* (28), 9902-9905. DOI 10.1021/ja505302q.

33. Ramirez, P.; Portillo, S.; Cervera, J.; Bisquert, J.; Mafe, S., *Phys. Rev. E* **2024,** *109* (4), 044803. DOI 10.1103/PhysRevE.109.044803.

34. Haynes, W. M., *CRC Handbook of Chemistry and Physics*. CRC Press: **2016**.

35. Powell, M. R.; Sullivan, M.; Vlassiouk, I.; Constantin, D.; Sudre, O.; Martens, C. C.; Eisenberg, R. S.; Siwy, Z. S., *Nat. Nanotechnol.* **2008,** *3* (1), 51-57. DOI 10.1038/nnano.2007.420.

36. Laucirica, G.; Toimil-Molares, M. E.; Marmisollé, W. A.; Azzaroni, O., *ACS Appl. Mater. Interfaces* **2024,** *16* (43), 58818-58826. DOI 10.1021/acsami.4c11522.

37. Tsutsui, M.; Hsu, W.-L.; Hsu, C.; Garoli, D.; Weng, S.; Daiguji, H.; Kawai, T., *Nat. Commun.* **2025,** *16* (1), 1089. DOI 10.1038/s41467-025-56052-0.

38. Liu, R.; Liu, Z.; Li, J.; Qiu, Y., *Biomicrofluidics* **2024,** *18* (2), 024103. DOI 10.1063/5.0203512.
39. Ju, B.-F.; Chen, Y.-L.; Ge, Y., *Rev. Sci. Instrum.* **2011,** *82* (1), 013707. DOI 10.1063/1.3529880.
40. Ma, T.; Janot, J.-M.; Balme, S., *Small Methods* **2020,** *4* (9), 2000366. DOI 10.1002/smtd.202000366.
41. Qiu, Y.; Vlassiouk, I.; Chen, Y.; Siwy, Z. S., *Anal. Chem.* **2016,** *88* (9), 4917-4925. DOI 10.1021/acs.analchem.6b00796.
42. Ma, L.; Zhang, H.; Ai, B.; Zhuang, J.; Du, G.; Qiu, Y., *J. Chem. Phys.* **2025,** *162* (9), 094704. DOI 10.1063/5.0253840.
43. Cheng, L.-J.; Guo, L. J., *Chem. Soc. Rev.* **2010,** *39* (3), 923-938. DOI 10.1039/B822554K.
44. Zhang, H.; Ma, L.; Zhang, C.; Qiu, Y., *Langmuir* **2024,** *40* (41), 21866-21875. DOI 10.1021/acs.langmuir.4c03204.
45. White, H. S.; Bund, A., *Langmuir* **2008,** *24* (5), 2212-2218. DOI 10.1021/la702955k.
46. Wei, J.; Du, G.; Guo, J.; Li, Y.; Liu, W.; Yao, H.; Zhao, J.; Wu, R.; Chen, H.; Ponomarov, A., *Nucl. Instr. Meth. Phys. Res. B* **2017,** *404*, 219-223. DOI 10.1016/j.nimb.2016.12.015.
47. Duleba, D.; Dutta, P.; Denuga, S.; Johnson, R. P., *ACS Meas. Sci. Au* **2022,** *2* (3), 271-277. DOI 10.1021/acsmeasuresciau.1c00062.
48. Lin, K.; Lin, C.-Y.; Polster, J. W.; Chen, Y.; Siwy, Z. S., *J. Am. Chem. Soc.* **2020,** *142* (6), 2925-2934. DOI 10.1021/jacs.9b11537.
49. Luo, L.; Holden, D. A.; Lan, W.-J.; White, H. S., *ACS Nano* **2012,** *6* (7), 6507-6514. DOI 10.1021/nn3023409.
50. Lin, C.-Y.; Wong, P.-H.; Wang, P.-H.; Siwy, Z. S.; Yeh, L.-H., *ACS Appl. Mater. Interfaces* **2020,** *12* (2), 3198-3204. DOI 10.1021/acsami.9b18524.
51. Song, F.; An, X.; Ma, L.; Zhuang, J.; Qiu, Y., *Langmuir* **2022,** *38* (42), 12935-12943. DOI 10.1021/acs.langmuir.2c02060.
52. Shi, D.; Wang, W.; Liang, Y.; Duan, L.; Du, G.; Xie, Y., *Nano Lett.* **2023,** *23* (24), 11662-11668. DOI 10.1021/acs.nanolett.3c03518.